\documentclass{emulateapj}
\usepackage{natbib,epsfig}

\newcommand{\rmd}{{\rm d}}

\begin{document}

\title[
Impact of intrinsic alignments on photometric redshift requirements]
{Dark energy constraints from cosmic shear power spectra:
impact of intrinsic alignments on photometric redshift requirements}

\author{Sarah Bridle$^{1,}$ and Lindsay King$^{2,}$}
\affil{$^{1}$ Department of Physics \& Astronomy, University College London, London, WC1E 6BT, U.K.\\
$^{2}$ Institute of Astronomy, Cambridge University, Madingley Road, Cambridge, CB3 0HA, U.K.
}

\begin{abstract}
Cosmic shear constrains cosmology by exploiting the apparent
alignments of pairs of galaxies due to gravitational lensing
by intervening mass clumps.
However galaxies may become (intrinsically) aligned with each other,
and with nearby mass clumps, during their formation.
This effect needs to be disentangled from the cosmic shear signal
to place constraints on cosmology.
We use the linear intrinsic alignment model as a base and compare it to
an alternative model and data. If intrinsic alignments are ignored then the dark
energy equation of state is biased by $\sim50$ per cent.
We examine how the number of tomographic redshift bins
affects uncertainties on cosmological parameters and find that
when intrinsic alignments are included two or more times as
many bins are required to obtain 80 per cent of the available information.
We investigate how the degradation in the dark energy figure of merit
depends on the photometric redshift scatter.
Previous studies have shown that lensing does not place stringent
requirements on the photometric redshift uncertainty, so long
as the uncertainty is well known.
However, if intrinsic alignments are included the
requirements become a factor of three tighter. These results are quite insensitive to the fraction of
catastrophic outliers, assuming that this fraction is well known.
We show the effect of uncertainties in photometric redshift bias
and scatter. Finally we quantify how priors on the intrinsic alignment model would
improve dark energy constraints.
\end{abstract}
%\begin{keywords}
%large-scale structure of Universe -- cosmological parameters -- surveys
%\end{keywords}

\maketitle

\section{Introduction}
\renewcommand{\thefootnote}{\fnsymbol{footnote}}
\setcounter{footnote}{1}
\footnotetext{E-mail: sarah.bridle@ucl.ac.uk}
The tidal gravitational field of density inhomogeneities in
the universe distorts the images of distant galaxies. This so-called
`cosmic shear' results in correlations in the observed ellipticities of
the distant galaxies, a signal which depends upon
the geometry of the universe and the matter
power spectrum \citep{Blandford91,Miraldaescude91,kaiser92}. The
first detections of cosmic shear made in 2000
\citep{baconre00,kaiserwl00,vanwaerbekeea00,wittmanea00}
demonstrated its value as a cosmological tool. Future generations of
multi-color imaging surveys will cover thousands of square degrees and have
the potential to measure the dark matter power spectrum
with unprecedented precision
in three dimensions at low redshift which is not possible with the CMB.
This is important because dark energy only starts to dominate at low redshift.
Ultimately, cosmic shear data can improve constraints on
the dark energy equation of state from the
CMB by over an order of
magnitude \cite{hu02}.

The great promise
and exactitude
of cosmic shear has necessitated the design of
instruments expressly geared towards measurement of the tiny
lensing-induced distortions. It has also motivated improvements in techniques
to account for changes in galaxy shapes due to the atmosphere and
telescope optics. Furthermore it has prompted careful consideration of any
potential cosmological contaminants of the cosmic shear signal.
Intrinsic alignments of galaxies are a potential contaminant
and fall into two categories. The first is
intrinsic-intrinsic galaxy alignments (II
correlations), which may arise during the galaxy formation process
since neighboring galaxies reside in a similar tidal field (e.g.
\cite{crittendennpt01}).
The second, related, effect
is a cross-term between intrinsic ellipticity and cosmic shear (GI correlations,
\cite{hiratas04}), whereby the
intrinsic shape of a galaxy is correlated with the surrounding
density field, which in turn contributes to the lensing distortion
of more distant galaxies. The net effect of this is an induced
anti-correlation between galaxy ellipticities, leading to a
suppression of the total measured signal.

Intrinsic galaxy alignment has been subject to numerous analytic,
numerical and observational studies [e.g.
\cite{croftm00,HRH,crittendennpt01,Mackeywk02,Jing02,heymansea04,hirataea04,bridlea07}].
\cite{brownthd02} and \cite{heymansea04} detected an II signal
in the SuperCOSMOS data.
\cite{mandelbaumhisb06} used in excess of a quarter of a million
spectroscopic galaxies from SDSS to obtain constraints on intrinsic
alignments, with no detection of an II signal.
Since II
correlations only operate when galaxies are physically close, this
contaminant is relatively straightforward to deal with by
down-weighting or removing those pairs
if redshifts are well known \citep{kings02,heymansh03,takadaw04} or by fitting parameterized
models relying on the different behavior of lensing and II in
redshift space \citep{kings03}.

The first observational detection of a large-scale density-galaxy
ellipticity correlation was made by \cite{mandelbaumhisb06}, using the
same SDSS sample noted above; their detection is robust and is
present on scales up to 60\,$h^{-1}$ Mpc.  They estimate that the
amplitude of the GI correlation could cause existing deep surveys to
underestimate the linear amplitude of fluctuations by as much as
$\sim20\%$. The GI signal is dominated by the brightest galaxies,
possibly due to these being BCGs (brightest cluster galaxies)
aligned with the cluster ellipticity. \cite{hirataea07} perform a
more detailed characterization of this effect,
including a higher redshift sample of LRGs (luminous red galaxies).
They estimate that results on $\sigma_8$ from future cosmic shear surveys
may be biased by around 10 per cent.
Using N-body simulations,
\cite{heymanswhea06} estimate that in a survey with median depth
$z_{m}\sim 1$, the GI signal can contribute up to 10\% of the
lensing signal on scales up to 20 arcmin.

As noted by \cite{hiratas04},  unlike II correlations, GI
correlations are not localized in redshift, and hence their removal
is more complex. Given that the GI term has been observationally
detected
we cannot ignore it.
\cite{hiratas04} suggested that the dependence of the GI signal on
redshift could be used to distinguish it from cosmic shear.
\cite{king05} showed that projecting the signal into a set of
template functions would enable the lensing, II and GI signals to be
isolated, again harnessing their different dependence on redshift.

In this paper we aim to remove the intrinsic alignment contamination
of cosmic shear by simultaneously measuring cosmological and intrinsic
alignment parameters from shear power spectra.
The increased number of fitted parameters may degrade constraints
on cosmological parameters.
Here we focus on the requirements this places on photometric
redshift quality.

The planned future imaging surveys will observe at least hundreds of
millions of galaxies. It will not be possible to obtain spectroscopic
redshifts for them all.
Therefore it is planned to rely on `photometric redshifts':
estimates calculated from galaxy luminosities in several observing bands.
The resulting redshift accuracy is critically dependent on the number
of observing bands used.
Equally important for measurement of cosmological parameters is
the existence of a representative sample with spectroscopic
redshifts which allows the redshift accuracy to be quantified.
It is important to plan now for these future observations since
decisions are being made about which observing bands to use.
Furthermore it may be necessary to mount a dedicated observing
campaign if more spectroscopic redshifts are needed than are
currently planned.

In practice additional information may be included in the fit to
help offset the degradation
due to using many free parameters to encapsulate the intrinsic alignment model.
This may come from measurements such as those by \cite{mandelbaumhisb06} and
\cite{hirataea07}.
Any additional information on at least the functional form of
the intrinsic alignment contributions from theory would also be a great help.
This information could be included by applying priors to the intrinsic
alignment power spectra in the simultaneous analysis.
We investigate the size of prior required to improve constraints relative
to the self-calibration regime.

In section~\ref{sec:ia} we introduce our fiducial intrinsic alignment model.
Section~\ref{sec:de} shows how dark energy constraints are affected by
varying parameters within this model.
In section~\ref{sec:photoz} we show the requirements this places on
photometric redshift quality.
Finally in section~\ref{sec:priors} we consider the effect of priors on
photometric redshift and intrinsic alignment parameters.

\section{Intrinsic Alignment Model}
\label{sec:ia}

In this paper we attempt to simultaneously measure the intrinsic
alignment and cosmological model from cosmic shear power spectra.
We therefore have to parameterize the sources of the intrinsic alignments
in some sufficiently flexible yet physically reasonable way.
We also need to define a fiducial model for intrinsic alignments
to put into our simulation of the future observations.

We assume a catalogue of source galaxy positions and shears which is divided up into
a number of bins in redshift, with a number per unit comoving distance
$n_{i}(\chi)$ for bin $i$.
For convenience this is normalized as $\int n_{i}(\chi) d\chi = 1$.
The lensing efficiency function for lensing a mass at redshift $z_d$
for source redshift bin $i$ may then
be written as
\begin{equation}
q_i(\chi_d) =
{3\over 2}\Omega_m
\frac{H_0^2}{c^2}
(1+z_d)
\int_0^\infty
\!\!\!\!\!
n_i(\chi_s)
\frac{(\chi_{s}-\chi_d) \chi_{d}}{\chi_s}
d\chi_s
\label{eq:wint}
\end{equation}
where $\chi_d$ and $\chi_s$ are the comoving distances from the observer
to the deflector and source respectively.
We assume a flat universe throughout.
$\Omega_m$ is the matter density parameter, $H_{0}$ is the Hubble
constant
and $c$ is the speed of light.

As detailed in \cite{hiratas04}
the shear power spectra between redshift bins $i$ and $j$
come from the sum of three terms
\begin{equation}
C_{\ell(ij)} = C^{GG}_{\ell (ij)} +
C^{II}_{\ell(ij)} + C^{GI}_{\ell(ij)}
\label{eq:cesum}
\end{equation}
where the first term is the usual gravitational lensing contribution,
the second term arises from the intrinsic alignments of physically close galaxies
and the final term arises from the intrinsic alignment of a galaxy with
a mass which lenses a more distant galaxy.

In this paper we consider only E mode power spectra, since the
lensing and intrinsic alignment contributions to the B mode power
spectra are both very small (for the model considered).

These contributions can be written in terms of underlying power spectra as
\begin{eqnarray}
C^{GG}_{\ell(ij)}
&=&\!\! \int_0^\infty \!{
q_i(\chi) q_j(\chi)
\over\chi^2}
P_\delta(k;\chi)
d\chi
\label{eq:powerspec_GG}
\\
C^{GI}_{\ell(ij)}
&=&\!\! \int_0^\infty \!{
(q_i(\chi) n_j(\chi) + n_i(\chi) q_j(\chi))
\over\chi^2}
P_{\delta,\tilde\gamma^I}(k;\chi)
d\chi
\label{eq:powerspec_GI}
\\
C^{II}_{\ell(ij)}
&=&\!\! \int_0^\infty \!{
n_i(\chi) n_j(\chi)
\over \chi^2}
P_{\tilde\gamma^I}(k;\chi)
d\chi
\label{eq:powerspec_II}
\end{eqnarray}
where $k={\ell/\chi}$.
The first power spectrum $P_\delta(k;\chi)$ is simply the (non-linear)
matter power spectrum at the deflector redshift.
For simplicity in this paper we use the \cite{peacockd96} method to
modify a~\cite{ma96} power spectrum (containing baryon supression but no wiggles).

The remaining two power spectra are not well known.
We use perturbations around the linear alignment model normalized
approximately to data.
We detail our assumptions and compare them to data below.

\subsection{Fiducial intrinsic-intrinsic term (II)}

\begin{figure}
\center
\epsfig{file=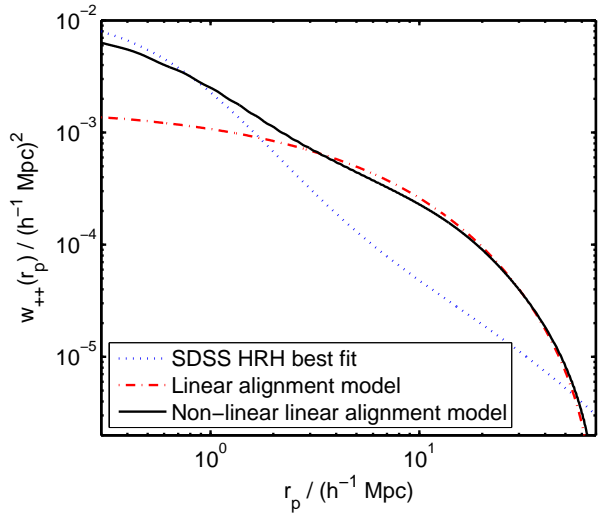,width=8cm}
\caption{
\emph{Dotted line}: The best fit to the SDSS data using the HRH* model.
\emph{Dot-dashed line}: The prediction from the linear alignment model
roughly normalized to~\cite{hirataea04}, who normalize to SuperCOSMOS.
\emph{Solid line}: The prediction from the linear alignment model
if a non-linear matter power spectrum is used in place of the linear
theory matter power spectrum.
The solid line is closer to the HRH* model and we use this model
throughout.
The error bars of \cite{mandelbaumhisb06} encompass all the models.}
\label{fig_fidii}
\end{figure}

The intrinsic-intrinsic term
describes the extent to which nearby galaxies are aligned with each other,
for example due to common tidal forces during their formation.
For this paper we simplify the linear alignment model presented in
\cite{catelankb01} and developed in~\cite{hirataea04} to use only the first term
of~\cite{hirataea04} Equation 16, written
\begin{equation}
P^{\rm lin}_{\tilde\gamma^I}(k) = {C_1^2\bar\rho^2\over\bar D^2}
P_\delta^{\rm lin}(k)
\label{eq:emode}
\end{equation}
where $C_1$ is a normalization constant,
and $P_\delta^{lin}(k)$ is the linear theory matter power spectrum.
$\bar D(z)\equiv (1+z)D(z)$ where $D(z)$ is the growth factor
normalized to unity at the present day.
$\bar\rho$ is the mean matter density of the Universe as a function
of redshift.
We estimate the value of $C_1$ by matching to the power spectra in
Figure 2 of~\cite{hirataea04} and estimate
$C_1=5\times10^{-14} (h^2 M_\odot / {\rm Mpc}^{-3})^{-2}$.
\cite{hirataea04} chose their $C_1$ by comparison with
SuperCOSMOS~\cite{brownthd02}.

The term we ignore is roughly an order of magnitude smaller than
the term we include.
This can be seen in~\cite{hirataea04} by examining the B-mode II power spectrum
which is of the same order as the term we ignore. Indeed from the figures
it can be seen that the B-mode II power spectrum is
typically an order of magnitude below the E-mode II power spectrum.

The linear alignment model might be a reasonable approximation on
large scales, but clearly makes a number of simplifying assumptions
on small scales. Inspired by the remark in \cite{hirataea07}
(Section 8)
we attempt to make a more realistic model by inserting the
full non-linear matter power spectrum $P_\delta(k)$ into~\ref{eq:emode}
\begin{equation}
P^{\rm nl}_{\tilde\gamma^I}(k) = {C_1^2\bar\rho^2\over\bar D^2}
P_\delta(k).
\label{eq:emode_nl}
\end{equation}
We refer to this throughout as the non-linear linear alignment model.
As noted in~\cite{hirataea07} this ``model" has no grounding in theory.
It might not even be closer to the truth:
galaxies may become more or less aligned due to non-linear interactions.
However we find this ``model" matches slightly better to other models and data
and therefore we use it as our default throughout.

The HRH* model for the intrinsic alignment correlation function was proposed by
\cite{hrh00} and improved by \cite{heymansbhmtw04}.
This model predicts the average intrinsic alignment on the sky
between two galaxies, as a function of the three dimensional separation
between the galaxies.
\cite{heymansbhmtw04} and \cite{mandelbaumhisb06} placed constraints
on the amplitude of this model using COMBO-17 and SDSS data
respectively.

These correlation functions may be projected along the line of sight to
produce quantities, $w_{++}$ and $w_{\times\times}$, which are
closely related to the II power spectrum in Eq.~\ref{eq:powerspec_II} by
\begin{equation}
P_{\tilde\gamma^I}(k) = 2\pi \int
w_{++}(r_p) J_{\,0}(kr_p)\,\,
r_p\,\rmd r_p
\label{eq:je}
\end{equation}
where $r_p$ is the separation between pairs of galaxies in the plane of the sky, and
throughout $J_{\,n}(F)$ are Bessel functions of the first kind.
Here $w_{++}$ is the correlation of galaxy shapes using the component
of shear aligned along the line joining the two galaxies.
There is another correlation function $w_{\times\times}$ which
comes from the shear components at 45 degrees to the line joining
the two galaxies.
In the above we have followed HRH* and assumed that these two
correlation functions are the same ($w_{++}=w_{\times\times}$)
and thus simplified~\cite{hirataea04} Equation 10 to write Eq.~\ref{eq:je}.

The HRH* model prediction for $w_{++}$ is given by
\begin{equation}
w_{++}(r_p) = \frac{A}{8{\cal R}^2}\int
\left[ 1+ \left(\frac{r}{r_0}\right)^{-\gamma_{gg}}
\right] \frac{1}{1+(r/B)^2}\, \rmd r_\parallel
\label{eq:hrhstar}
\end{equation}
where $\cal R$ is the factor to convert measured ellipticity to shear,
$r_\parallel$ is the separation between galaxy pairs along the line
of sight,
and $r=\sqrt{r_\parallel^2+r_p^2}$.
The galaxy clustering correlation function is described by
$r_0$ and $\gamma_{gg}$,
and $A$ and $B$ are free parameters of the HRH* model.

\cite{mandelbaumhisb06} follow \citet{heymansbhmtw04}
by fitting $A$ to data and using $\gamma_{gg}=1.8$, $B=1\,h^{-1}\,$Mpc, and
$r_0=5.25\,h^{-1}\,$Mpc. For their method ${\cal R}\sim0.87$.
We calculate $w_{++}$ using these parameters and the best fit to
SDSS L3-L6 of $A=1.8$.
We follow Section 4.3 of \cite{mandelbaumhisb06} by using a
range of integration of $0<r_\parallel/(h^{-1} {\rm Mpc})<60 h^{-1}$ Mpc.
The result is shown by the dotted line in  Fig.~\ref{fig_fidii}.
Our x axis range covers the range of data used for the fit.

We use the linear alignment model to predict $w_{++}$ by inverting
Eq.~\ref{eq:je}:
\begin{equation}
w_{++}(r_p) = \frac{1}{2\pi} \int
P_{\tilde\gamma^I}(k)
 J_{\,0}(kr_p)\,\, k \,\rmd k.
\label{eq:je_inverse}
\end{equation}
We calculate this at a redshift of $z=0.12$ to match the mean redshift
of the SDSS sample.
This is shown in Fig.~\ref{fig_fidii} by the dot-dashed line,
which uses the linear theory matter power spectrum (Eq.~\ref{eq:emode}
in Eq.~\ref{eq:je_inverse}).
The solid line shows the result when the non-linear theory
matter power spectrum is used (Eq.~\ref{eq:emode_nl} in Eq.~\ref{eq:je_inverse}).

It can be seen that the linear alignment model predictions have
a similar amplitude to the HRH* fit. This is not surprising
since the amplitude of the linear alignment model was taken
from~\cite{hirataea04} who made a fit to the \citet{heymansbhmtw04} data
which is not far from the \cite{mandelbaumhisb06} result.
The linear alignment model is lower than the HRH* model on small
scales and larger on scales $3 h^{-1} {\rm Mpc} <r_p <60 h^{-1}\,$ Mpc.
However the HRH* model was developed for small
scale correlations therefore the discrepancy with the linear alignment
model on large scales is not a concern (it was based on an approximation
to the density correlation function that only applies on small scales).
Both the linear alignment model and the HRH* model fit comfortably
within the error bars of
\cite{mandelbaumhisb06}.
We note that the non-linear linear alignment model is closer to the
HRH* model.

\subsection{Fiducial shear-intrinsic term(GI)}

\begin{figure}
\center
\epsfig{file=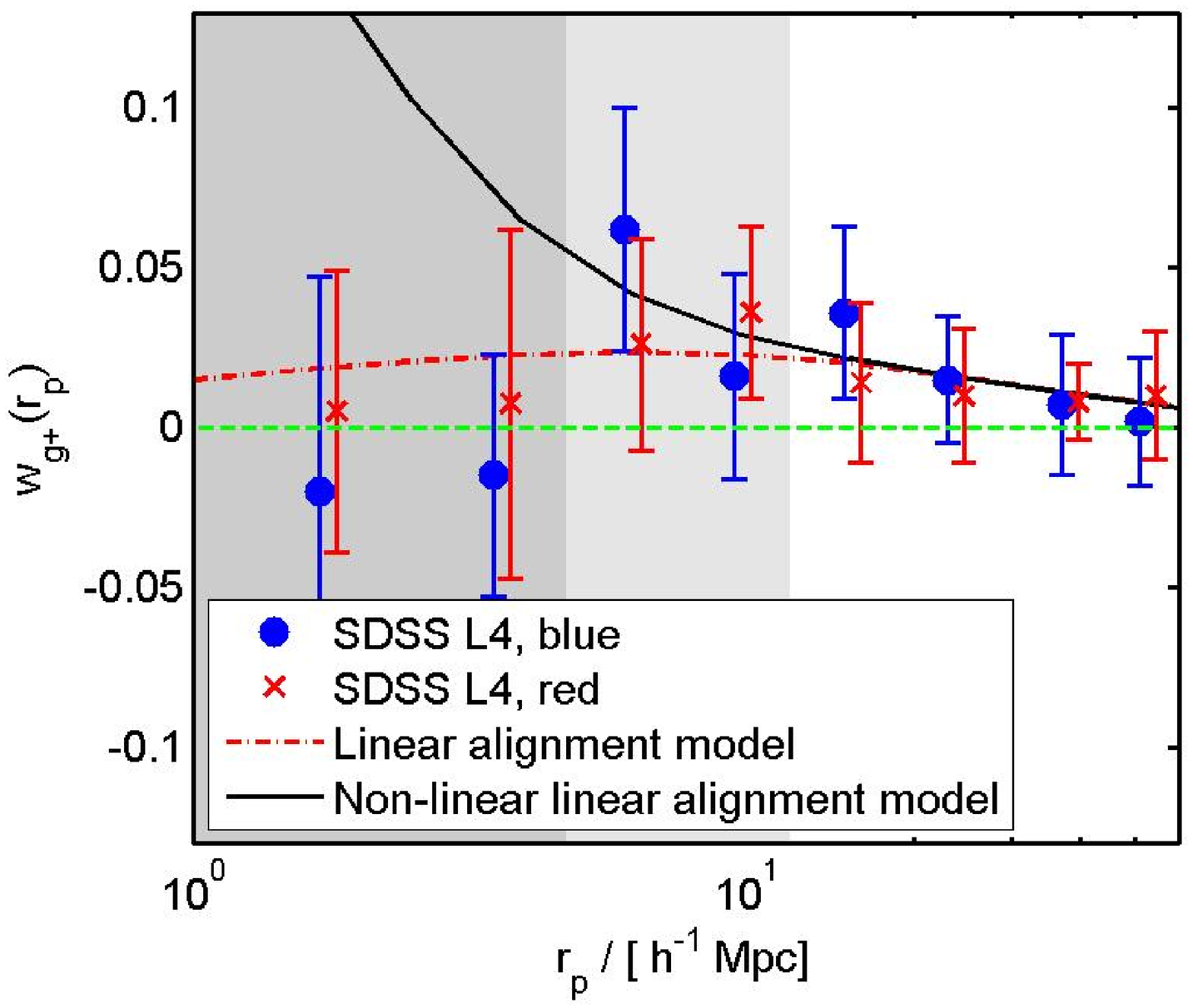,width=8cm}\\
\epsfig{file=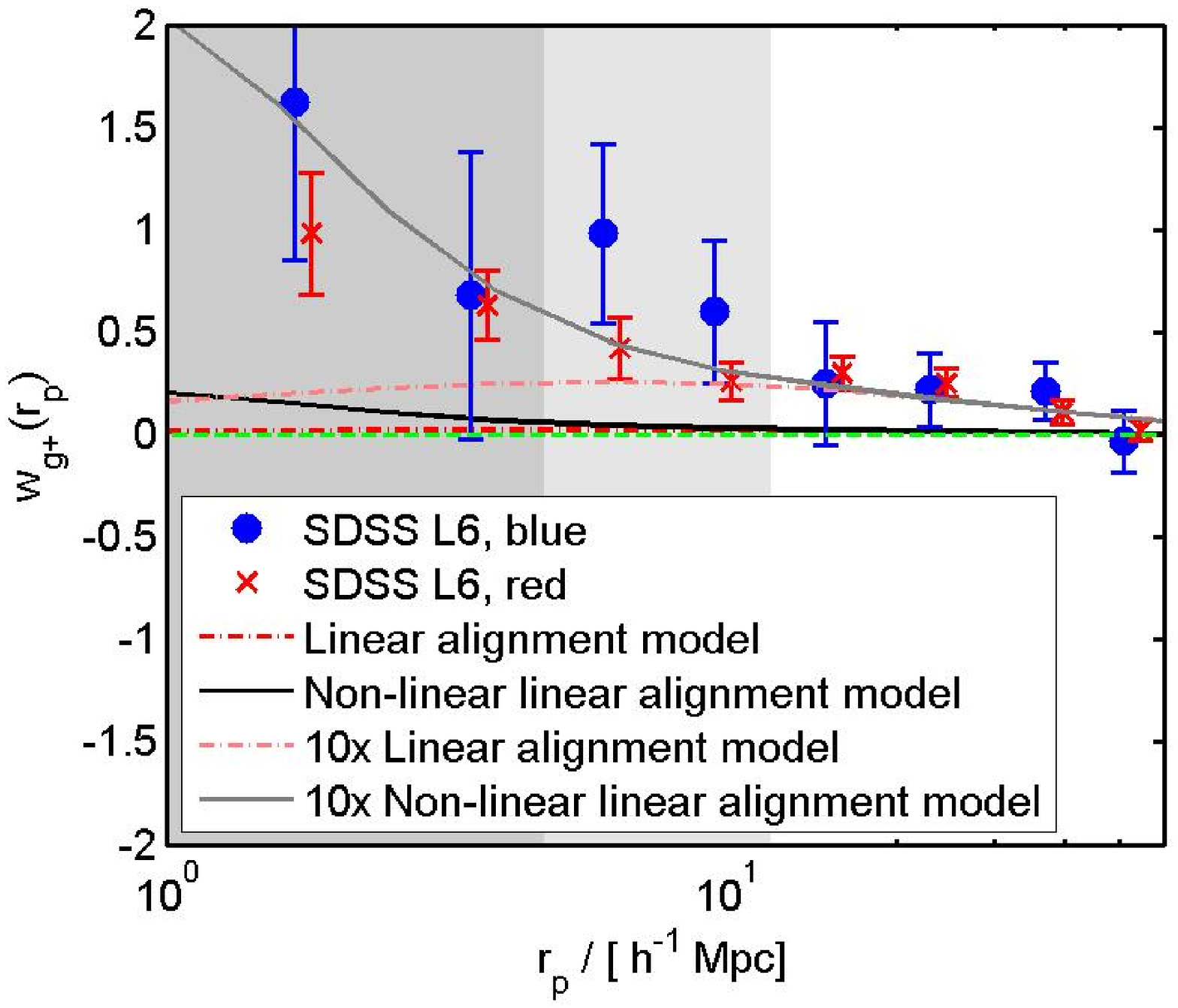,width=8cm}
\caption{
The dot-dashed line shows the linear alignment model.
The solid line shows our fiducial model, which uses the non-linear
matter power spectrum in the linear alignment model.
Data points taken from the analysis of SDSS in Figure 1 of~\cite{hirataea07}.
Circles are for blue galaxies and crosses for red galaxies.
Note that only $r_p>11.9 h^{-1}$ Mpc ($r_p>4.7 h^{-1}$ Mpc) were used in their
most (least) conservative fits (grey shading).
\emph{Upper} L4, the largest sample.
\emph{Lower} L6, the brightest sample. The upper two lines show the
linear alignment model predictions multiplied by a factor of ten.
}
\label{fig_fidgi}
\end{figure}

\begin{figure*}[t]
\center
\epsfig{file=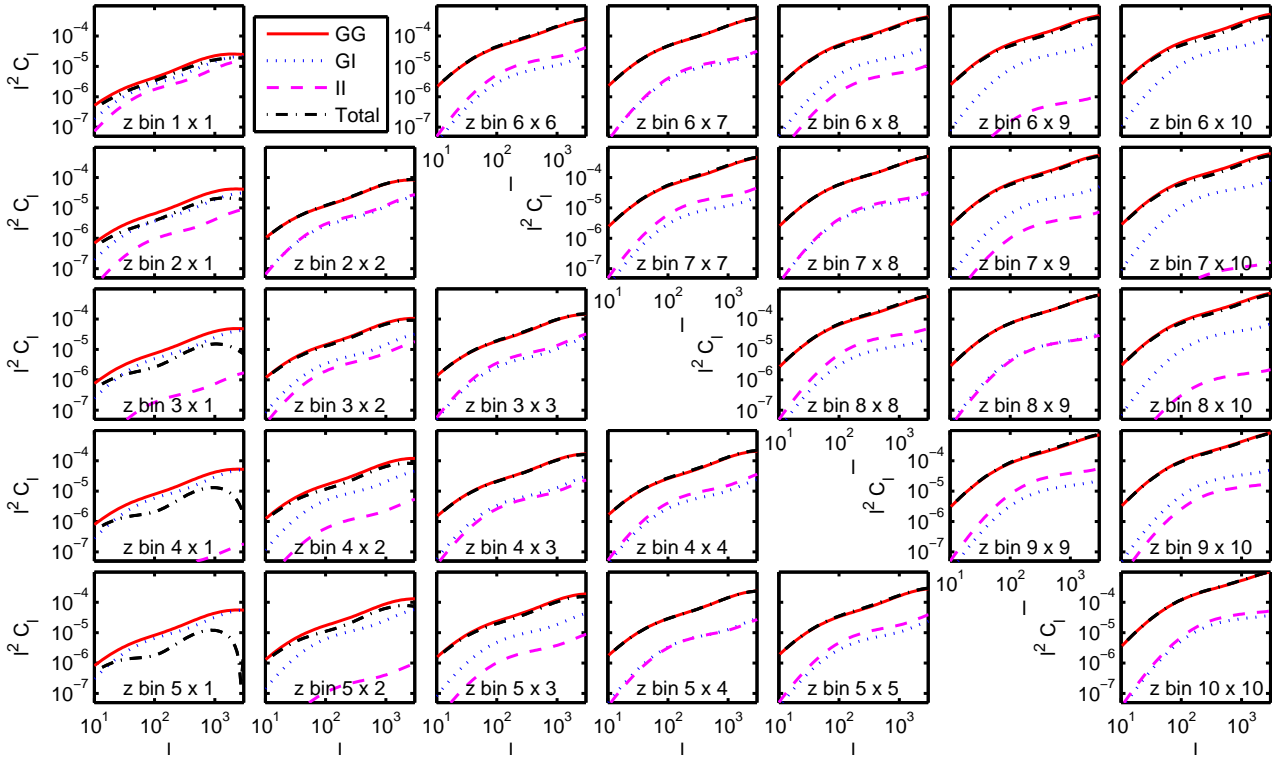,width=17cm}
\caption{A selection of shear cross power spectra for our fiducial
survey divided into ten tomographic redshift bins.
We assume a photometric redshift scatter of $\delta_z = 0.05 (1+z)$
and zero catastrophic outliers.
\emph{Solid line}: lensing shear (GG) term.
\emph{Dashed line}: intrinsic shear (II) term.
\emph{Dotted line}: shear-intrinsic (GI) cross term (absolute values shown).
\emph{Dot-dashed line}: total.
Galaxies are divided into tomographic redshift bins with equal numbers
of galaxies in each bin. The median redshifts of the tomographic
redshift bins 1 to 10 are
$(0.30, 0.49, 0.62, 0.73, 0.84, 0.96, 1.1, 1.2, 1.4, 1.9)$ respectively.
}
\label{fig_clbin_fid}
\end{figure*}

As shown in~\cite{hirataea04},
the linear alignment model predicts for the GI underlying power spectrum
\begin{equation}
P^{\rm lin}_{\delta,\tilde\gamma^I}(k) = -{C_1\bar\rho\over\bar D} P_\delta^{\rm lin}(k).
\label{eq:cross}
\end{equation}
where $C_1$ is the same number as in Eq.~\ref{eq:emode}.
We also define
\begin{equation}
P^{\rm nl}_{\delta,\tilde\gamma^I}(k) = -{C_1\bar\rho\over\bar D} P_\delta(k)
\label{eq:cross_nl}
\end{equation}
where $P_\delta$ is the full non-linear matter power spectrum
which may help to encapsulate non-linear effects on smaller scales.
For example it was shown in~\cite{bridlea07} that even if there were no
tidal interactions between galaxies, there will still be a
shear-intrinsic correlation on small scales if galaxies are aligned with their
own dark matter halos.

We compare this to the SDSS results by \cite{hirataea07},
who measure the real space galaxy-shape correlation function
projected along the line of sight $w_{g+}$
which is defined by
\begin{equation}
w_{g+}(r_p)=\int \xi_{g+}(r_p,\Pi) {\rm d} \Pi
\end{equation}
where $\xi_{g+}(r_p,\Pi) = \langle \gamma_{+} \rangle_{(r_p,\Pi)}$ is the radial
shear of one galaxy relative to another, averaged over all pairs of
galaxies of a separation $r_p$ on the sky and $\Pi$ along the line of
sight (see \cite{hirataea07} equation 10 for the estimator they use).

The \cite{hirataea07} results for luminosity bins L4 and L6 are shown in Fig.~\ref{fig_fidgi}.
We show their results for L4 because it has the largest number of galaxies and therefore
will be most typical of future cosmic shear surveys.
We show results for L6 (which contains the most luminous galaxies) because there is a detection of the
signal
in this band and we would like to compare the shape of the linear
alignment model prediction with the observations.

We can more easily predict theoretically the \emph{mass}-shape correlation
function $w_{\delta+}$ which
can be related to the corresponding power spectrum by
\begin{equation}
w_{\delta+}(r_p) = -\frac{1}{2\pi} \int
P_{\delta,\tilde\gamma^I}(k)
 J_{\,2}(kr_p)\,\, k \,\rmd k
\label{eq:gi_inverse}
\end{equation}
\citep[][Equation 23]{hirataea07}.
\cite{hirataea07} convert from $w_{\delta+}$ to $w_{g+}$ by multiplying
by the galaxy-mass bias of the sample which we take from their Table 5.
We use the effective redshift values from this table and the
same cosmological model as~\cite{hirataea07} to calculate
$P_{\delta,\tilde\gamma^I}$ and convert it to $w_{g+}$
shown in Fig.~\ref{fig_fidgi}.

The L4 data points fit very nicely over the fiducial linear alignment
model predictions.
Note that the dark (light) grey shaded regions block over the data
points that~\cite{hirataea07} did not use for their most (least) conservative
cuts when fitting models.
It is not trivial that these data agree with the model.
The amplitude $C_1$ in the linear alignment model was chosen
to match observations of the II power spectrum.
The agreement with the GI data is some support for the linear alignment model
which links the two.

The L6 data points are much above our fiducial linear
alignment model predictions.
This is because we are aiming to use parameters correct for
the majority of the population, whereas L6 is a very bright
sample.
We multiply the predictions by a factor of 10 to compare their
shapes with the data.
We see that there is a slightly better fit when using
the non-linear matter power spectrum in the linear alignment
model equation (Eq.~\ref{eq:cross_nl}).
It was effectively this comparison which originally led~\cite{hirataea07} to suggest using the
non-linear power spectrum.

On scales less than $4.7 h^-1$ Mpc the L4 results appear to agree
slightly better with the linear theory predictions than those
from the non-linear matter power spectrum. However the error bars
are large and these data points are not used in the model fitting
by Hirata et al because the bias is non-linear on these scales.
We expect the non-linear alignment model to be more difficult
to disentangle from the cosmic shear signal because it has a
similar shape, as a function of scale.
We show in Fig.~\ref{fig_fom_deltaz} the results for both models and find them to
be quite similar in their photometric redshift requirements.

\begin{figure*}[t]
\center
\epsfig{file=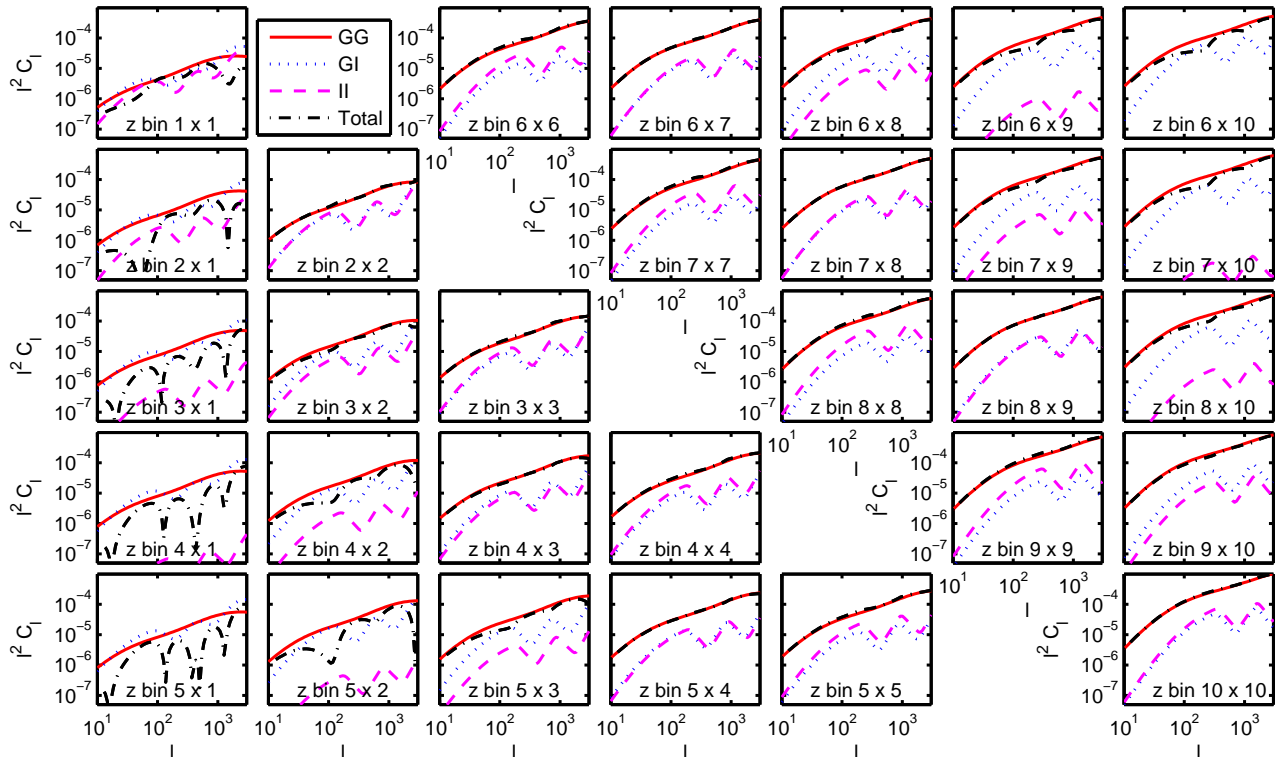,width=17cm}
\caption{
Illustration of the default spatial flexibility we allow in the
intrinsic alignment power spectra.
Lines are as in the previous figure but the intrinsic alignment
power spectra have been multiplied by a function with five bins in
k which are interpolated linearly in the log as detailed in the text.
We have arbitrarily set the bin values to
$B^X_{1 .. (n-1)}=\left(1,\, -1,\, 1,\, -1,\, 1\right)$
to show the freedom.
}
\label{fig_clbin_wiggles}
\end{figure*}

\subsection{Implications for our fiducial survey}

For the remainder of this paper we use as our fiducial cosmological
model the best fit to WMAP3~\citep{spergelea06}.
This has a normalized Hubble constant
$h \equiv H_0 / 100 {\rm km} s^{-1} {\rm Mpc}^{-1} =0.73$,
total matter density $\Omega_m=0.23$,
baryon density $\Omega_b h^2=0.022$,
fluctuation amplitude normalization $\sigma_8=0.74$
and
primordial fluctuation spectral index $n_s=0.95$.
We use fiducial dark energy parameters $w_0=-0.95$ and
$w_a=0$ where the dark energy equation of state as a function
of scale factor is given by $w(a)=w_0 + (1-a)w_a$
\citep{wa1,wa2}.
We use a maximum angular power spectrum multipole $l_{\rm max}=2\times10^4$
throughout.

Throughout we use a fiducial survey similar to
DUNE\footnote{The Dark UNiverse Explorer (DUNE) \url{http://www.dune-mission.net/}}
and LSST\footnote{The Large Synoptic Sky Survey  (LSST) \url{http://www.lsst.org}},
the `shallow' survey described in~\cite{amarar07}.
This covers 20,000 square degrees to a depth of $z_m=0.9$ with
$35$ galaxies per square arcminute which can be used for shear measurements.
As in~\cite{amarar07} we use the redshift distribution given by~\cite{smailef94}
using $\alpha=2$ and $\beta=1.5$.
By default we divide this survey up into twenty
redshift bins, binned to have
an equal number of source galaxies in each bin.
As discussed below we use $\delta_z$ to parameterize the photometric
redshift error, and by default use $\delta_z=0.05$.

To show the size of the intrinsic alignment contributions we plot in
Fig.~\ref{fig_clbin_fid} some of the power spectra
for a survey divided into 10 source redshift bins.
The GI and II contributions to the lensing power spectra are typically
about an order of magnitude lower than the GG contribution.
For illustration we show only the results for a high redshift
subset of the bins.
However, for the auto correlation of the lowest redshift bin the II signal
is the largest contribution to the total power spectrum.
When cross correlating the lowest redshift bin with the 2nd and 3rd lowest
bins, the GI term can become larger than the GG term and the total power
spectrum negative.

Despite the small size of the intrinsic alignment contribution, we must
note that the effect on the power spectra of the cosmological parameters, including
the dark energy equation of state,
can be very small and therefore these contributions are very important.
\cite{huterertbj06} give an approximate expression (equation 24 of their paper) for the
dependence of the lensing power spectra on cosmological parameters which includes
$C^{GG}_{\ell} \propto \sigma_8^{2.9} |w_0|^{0.31}$.
Therefore an increase in the fluctuation normalization parameter $\sigma_8$
by 1 per cent increases the power spectra by around 3 per cent.
A change in $w$ of 1 per cent (e.g. $w=-1$ to $w=-0.99$) typically
changes the lensing power spectrum by about $0.3$ per cent (decrease).
With future surveys we are aiming to measure the dark energy
equation of state to of order 1 per cent.

We simulate power spectra that include intrinsic alignment contributions
but find the best fit cosmological parameters that would be
obtained if intrinsic alignments were mistakenly ignored (GG).
The purpose is to assess the importance of bothering to consider
intrinsic alignments for future studies.
We make the simplistic assumption that changes in the cosmic shear
power spectra are linear in the cosmological parameters.
If this assumption were correct then Fisher matrix uncertainties
would also be correct, so it is a useful investigation to compare with
predicted cosmological parameter uncertainties.

Given the numbers above,
it is not surprising that if we fit to the lensing power
spectra mistakenly ignoring the GI and II contributions we mis-estimate
the fluctuation amplitude parameter $\sigma_8$ by $\sim 5 \%$,
in agreement with previous studies
(see the upper section of Table~1 for details).
From the numbers above we expect the effect on the dark energy
equation of state to be about a factor of 10 larger.
Indeed we find that the dark energy equation of state
is mis-estimated
by around 50 per cent for a non-tomographic (2D) analysis.

\begin{deluxetable}{clrrl}
\tablewidth{0pt}
\tablecaption{Percentage biases on cosmological parameters when
wrongly ignoring intrinsic alignments.}
\tablehead{
  \colhead{}
& \colhead{}
& \colhead{$\sigma_8$}
& \colhead{$w_0$}
& \colhead{$w_0$ (marginalised)}
}
\startdata
&GG+II &                     1.7\% &             $ -0.24$   & ~~~0.28 \\
2D&GG+GI &                    $-5.5\%$ &               0.78   & ~~~6.8 \\
&GG+II+GI &  $-3.8\%$    &  0.55   & ~~~0.15 \\
\hline
&GG+II &                      3.2\% &             2.1   & ~~~5.8 \\
3D&GG+GI &                    $-2.9\%$ &              0.03   & ~~~1.5\\
&GG+II+GI &  0.3\%    &  2.5  &  ~~~4.3\\
\enddata
\tablenotetext{}{
NOTE. $-$\emph{Upper section:} for a non-tomographic (2D) analysis.
\emph{Lower section:} for 10 bin tomographic analysis.
Columns show:
the percentage change in $\sigma_8$ when fixing all
other cosmological parameters;
the absolute change in $w_0$ when
fixing all other parameters;
and for $w_0$ when marginalising
over the other 6 cosmological parameters used in this paper.
The results can be quite different when marginalising over other
parameters due to movement along degeneracy directions to better
fit the intrinsic alignment signal.}
\end{deluxetable}
\bigskip 

This mis-estimate only applies if all cosmological parameters are known
except the dark energy equation of state.
In fact we find that if several cosmological parameters are fitted simultaneously
the mis-estimate becomes even worse,
as near-degeneracy directions are exploited in an attempt to match
the shape of the GG+GI+II power spectra using the GG spectra alone.
Therefore if our fiducial model is roughly correct it would be a disaster
to ignore intrinsic alignments in studies aiming to measure the dark
energy equation of state.

Table~\ref{tab:biases} shows that the final
overestimate of the dark energy equation of state is 15\% for a
2D non-tomographic analysis, and over 400\% for a 10 bin
tomography analysis. Note that these numbers are only indicative,
since they assume  that the likelihood surface is Gaussian with
a constant curvature, which  will not be a good approximation so
far from the fiducial model.

The simplest way we can hope to remove the II contribution is to use
only cross-power spectra and to ignore the auto-correlations.
Fig.~\ref{fig_clbin_fid} reminds us that more than just the auto-correlations may need
to be removed if the photometric redshifts are not perfect and many
redshift bins are used. For 10 photometric redshift bins and
$\delta_z=0.05$ the II contribution is non-negligible even when
cross correlating non-neighboring bins (e.g. $i=6, j=8$).
Further, due to the photometric redshift errors the GI
contribution is non-negligible even for auto-correlations.

\subsection{Parameterizing intrinsic alignment uncertainties}

On obtaining shear power spectra from new data we will attempt to estimate
cosmological parameters.
However, since the linear alignment model is approximate we
cannot rely on it being correct.
It seems unlikely that the galaxy formation process will one
day be understood so well that we can use it to perfectly
remove intrinsic alignments.
We can hope to measure the contaminant from analyses such as
those of~\cite{mandelbaumhisb06} and~\cite{hirataea07}.
However it will still be necessary to
use a parameterized model to fit to the data points, and
there will remain some uncertainty on the model parameters
due to the error bars on the data points.

For this paper we parameterize perturbations around the linear
alignment model and show the effect of fitting these free parameters
simultaneously with the cosmological parameters.
This follows the approach indicated in section 7 of the technical
appendix of~\cite{detf}.
Our most basic perturbation is to assume some unknown amplitude
and redshift dependence, which is different for each of II and GI:
\begin{equation}
P^{\rm base}_{X}(k;\chi) = A_{X}
\left(\frac{1+z}{1+z_0}\right)^{\gamma_X}
P^{\rm nl}_X(k;\chi)
\end{equation}
where $X$ is ${\tilde\gamma^I}$ or ${\delta,\tilde\gamma^I}$.

However this assumes that the scale dependence of the models
is well known, whereas we have seen that there is still great
uncertainty (e.g. compare the HRH model with the linear alignment
model in Fig.~\ref{fig_fidii}).
Further, the scale will likely deviate in a different way from the linear
alignment model at different redshifts.
In principle we have two unknown 2d functions
$Q_{\tilde\gamma^I}(k;\chi)$ and
$Q_{\delta,\tilde\gamma^I}(k;\chi)$
multiplying each
of the power spectra for the II and GI terms:
\begin{equation}
P^{\rm free}_{X}(k;\chi) =
Q_X(k;\chi)
P^{\rm base}_{X}(k;\chi).
\end{equation}

To parameterize the $Q_X$ using a finite number of parameters we
use $n$ bins in $k$ and $m$ bins in redshift and
linearly interpolate in the logs of $k$, $(1+z)$ and $Q_X$.
That is,
\begin{eqnarray}
\ln(Q_X(k,\chi)) =
K\,&Z&\,B^X_{ij} +
(1-K)\,Z\,B^X_{(i+1)j}
\nonumber \\
+ K\,(1-Z)\,B^X_{i(j+1)} &+& (1-K)\,(1-Z)\,B^X_{(i+1)(j+1)}
\label{eq:kzbins}
\end{eqnarray}
where
\begin{eqnarray}
K&=&\frac{\left[\ln(k)-\ln(k_i) \right]}{\left[ \ln(k_{i+1})-\ln(k_i) \right]}
\,\,\, k_i < k < k_{i+1}\\
Z&=&\frac{\left[\ln(1+z)-\ln(1+z_j) \right]}{\left[ \ln(1+z_{j+1})-\ln(1+z_j) \right]}
\,\,\, z_j < z < z_{j+1}
\label{eq:kzbins_KZ}
\end{eqnarray}
and $i$ runs from $0$ to $n$ and $j$ runs from $0$ to $m$.
Because we are interpolating on the log of a multiplicative function
setting all the $B^X_{ij}$ values to zero
makes the multiplicative function everywhere unity.

We use different free parameters $B^X_{ij}$ for II than for GI.
However we use the same number of bins $n$, $m$ for both of II and GI.
We set $k_0$ and $k_{n+1}$ to be at the edges of our $k$ integration ranges
($k_0=10^{-4} h {\rm Mpc}^{-1}$, $k_n=2\times 10^3 h {\rm Mpc}^{-1}$) and
fix $B_{0j}=B_{(n+1)j}=0$ for all $j$
i.e. the multiplicative function goes to unity at low and high $k$.
$k_1$ to $k_{n}$ are spaced linearly in the log in a smaller range in $k$.
We use $k_1=0.1 h {\rm Mpc}^{-1}$ and $k_{n}= 2 h {\rm Mpc}^{-1}$
because this spans from near the start of the non-linear regime to a point which
has less effect on our angular power spectra.

The basic effect on the angular power spectra is illustrated
in Fig.~\ref{fig_clbin_wiggles}.
Here we have used no evolution in redshift ($\gamma_X=0$, $m=1$)
and set $B_{1 .. n}=\left(1,\, -1,\, 1,\, -1,\, 1\right)$ to
highlight the degree of flexibility and the $\ell$ values corresponding
to the edges of the bins.

By default we use one bin in redshift ($m=1$) which means that
we assume the same power law
redshift evolution in each $k$ bin.
By default we use $n=5$ bins in $k$.
We explore the effect of these defaults below.
For numerical stability we apply a wide prior on each B value
of 10. We see in Section~\ref{sec:priors} that this is sufficiently wide that
it would give the same answer as a wider prior.

\section{Dark Energy Constraints}
\label{sec:de}

\begin{figure*}[t]
\center
\epsfig{file=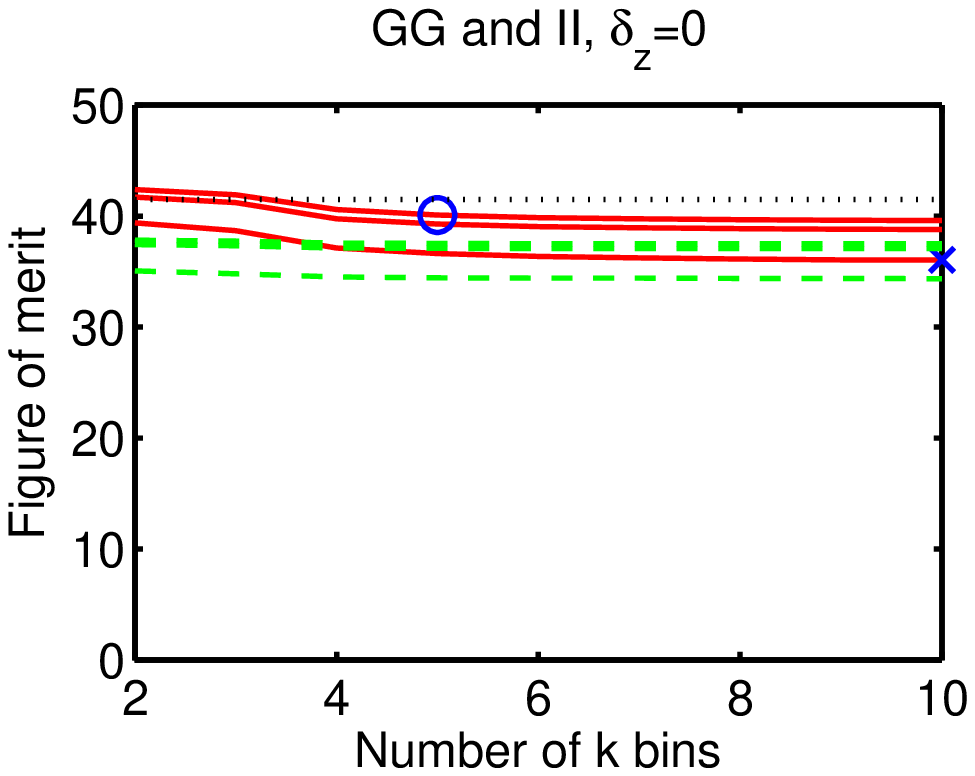,width=5.5cm}
\epsfig{file=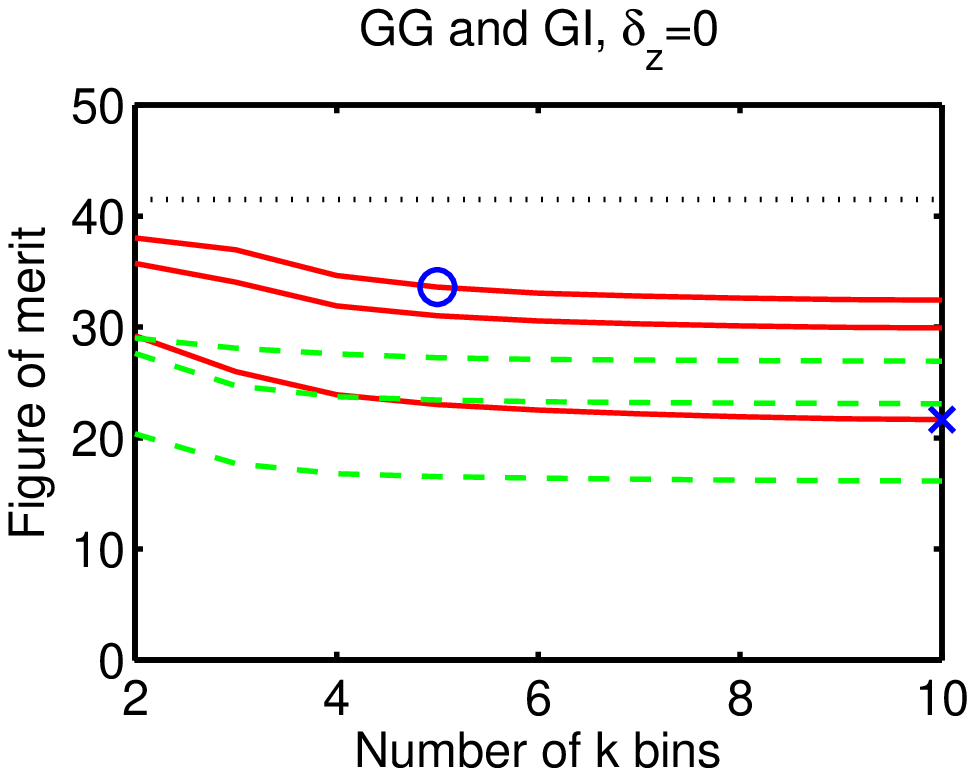,width=5.5cm}
\epsfig{file=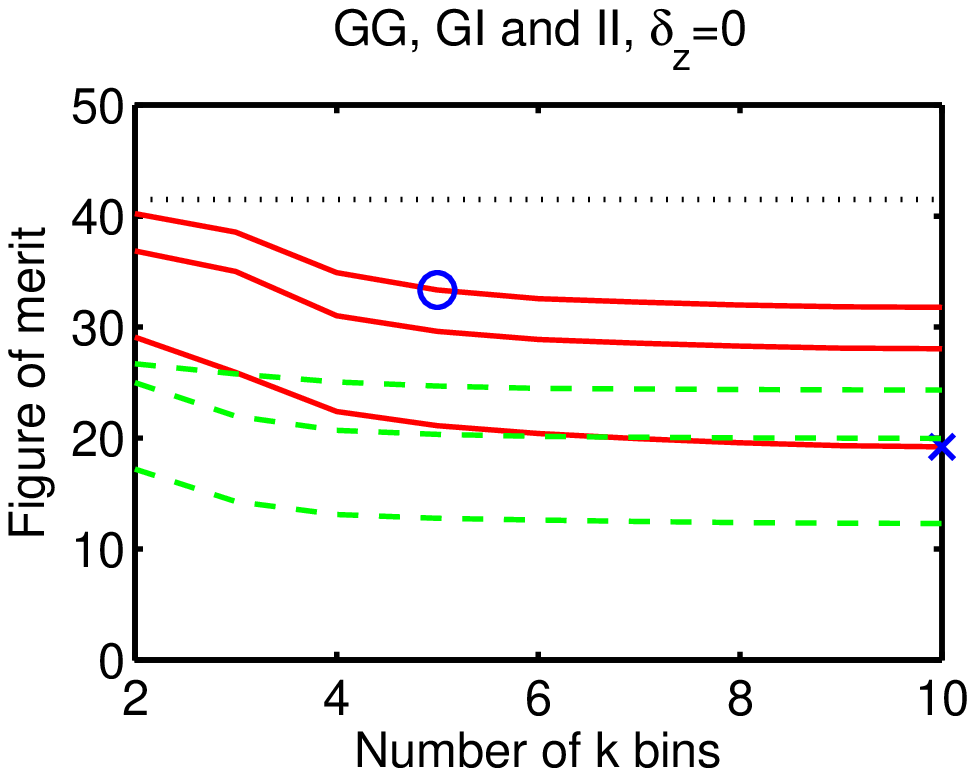,width=5.5cm}\\
\epsfig{file=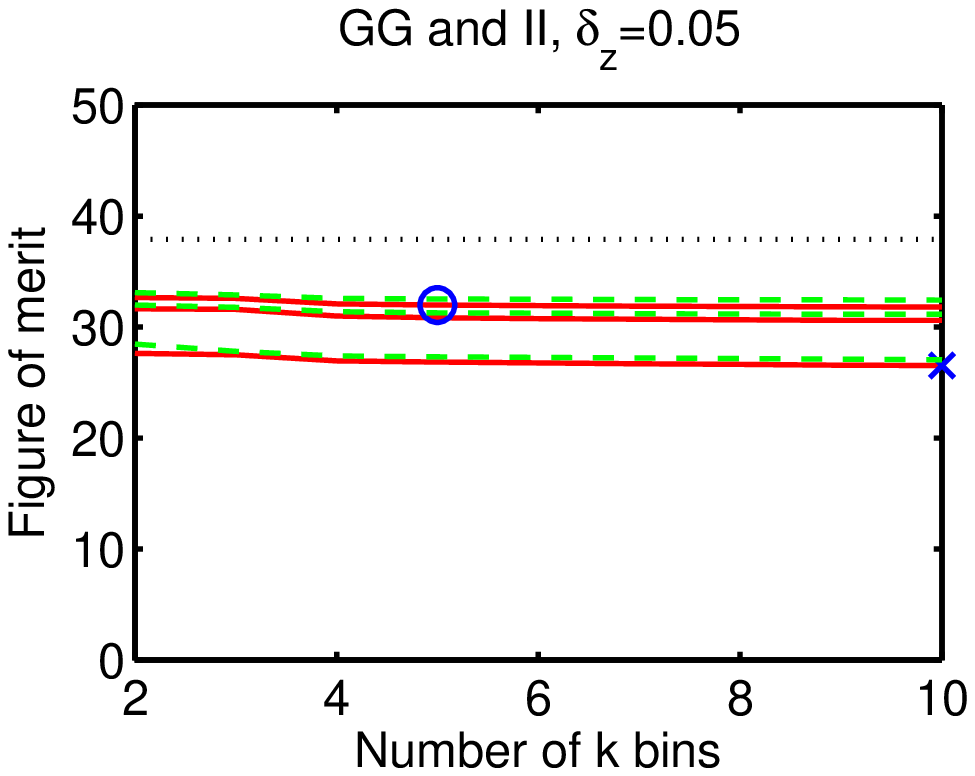,width=5.5cm}
\epsfig{file=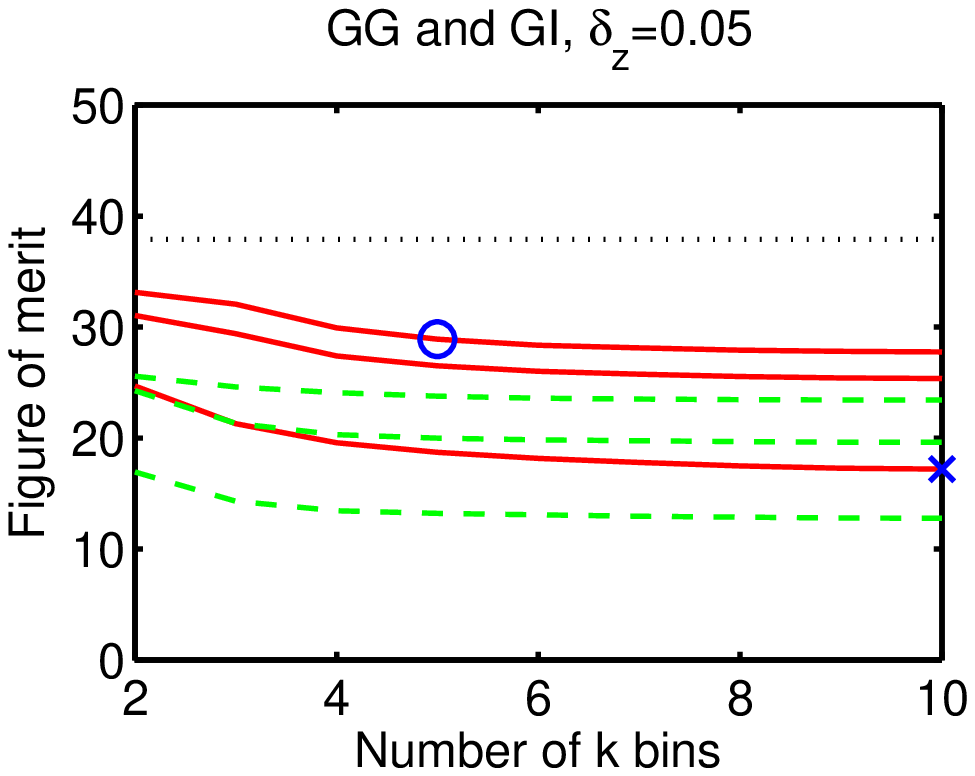,width=5.5cm}
\epsfig{file=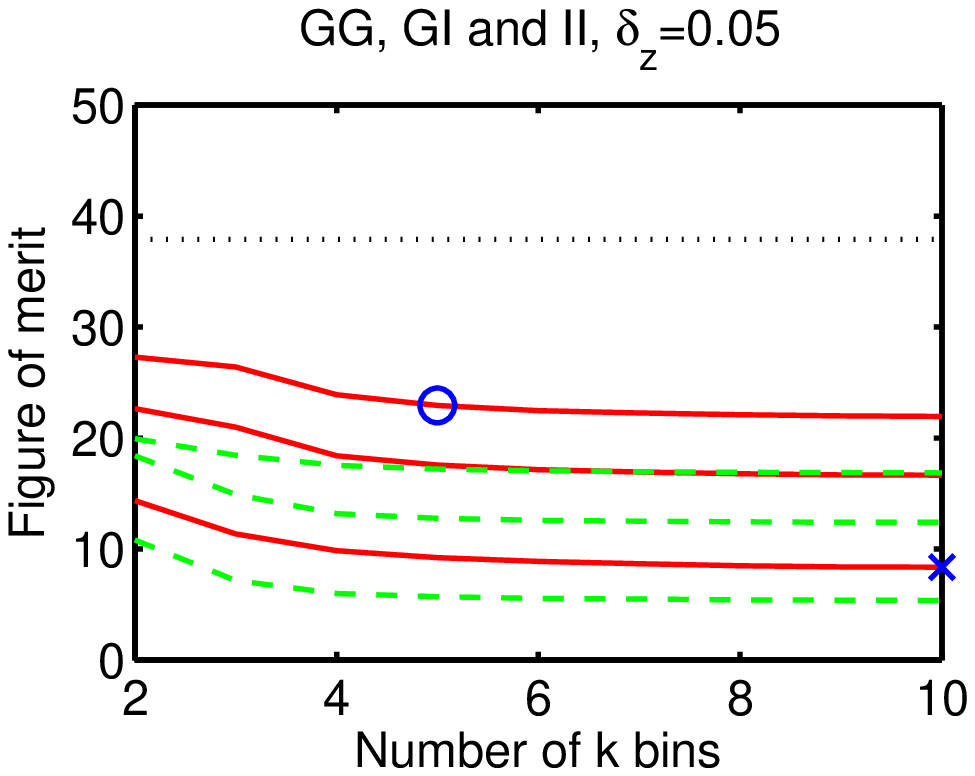,width=5.5cm}
\caption{
\emph{Top row:} Assuming perfect photometric redshifts ($\delta_z=0$).
\emph{Bottom row:} More realistic photometric redshifts ($\delta_z=0.05$).
\emph{Left column:} Including just the II terms.
\emph{Middle column:} Including just the GI terms.
\emph{Right column:} Including both II and GI terms.
Solid lines: figure of merit as a function of number of kbins.
Lines from top to bottom within one panel:
the number of bins in the redshift direction increases 1, 2, 5.
Dotted line: FoM for GG alone.
Dashed line: using the linear alignment model with the linear theory
matter power spectrum, instead of the non-linear theory matter power spectrum.
In every case a free amplitude parameter and unknown power law evolution
was marginalized over (allowed to be different for each of GI and II).
The default intrinsic alignment model we use for the remaining figures
is shown by a circle and our maximally flexible model is marked by a cross.
}
\label{fig_fom_nkbin}
\end{figure*}

We now consider the constraints on dark energy that can be placed
after marginalising over other cosmological and
intrinsic alignment parameters.
We give our results in terms of the Figure of Merit (FoM) advocated by
the Dark Energy Task Force report~\citep{detf}. This is the inverse of the
area of the 95 per cent contour in the $w_0$, $w_a$ plane.
We marginalise with flat priors over
$\Omega_m$, $h$, $\sigma_8$, $\Omega_b$ and $n_s$ assuming a flat Universe.
Note that the WMAP3 cosmological parameters give a smaller FoM compared
to most other sets of parameters, since they have lower values for
$\Omega_m$, $\sigma_8$ and $n_s$. For acceptable values of these parameters
the absolute value of the FoM can be a factor of two or more higher
for the survey parameters we assume. However this roughly scales the
FoM the same for all subsidiary parameter values we consider.

We use a Fisher matrix approach to quantify uncertainties.
We use the equations in~\cite{takadaj04_cospars}.
We add the GI and II power spectra to the GG power spectra
and insert this as the simulated total power spectra.
When calculating the Fisher matrix we need the covariance matrix
between all the power spectra.
For the cosmic variance part we use the sum of all the power spectra.
Because the GI term is negative this sum can in principle be zero.
Therefore we checked the effect of using only the lensing (GG) term
in the covariance matrix. This gives almost the same results, except
the figures of merit are slightly bigger, particularly for the GG+II
results due to the decreased error bars on ignoring the positive II contribution.

In principle the linear alignment model predictions depend on cosmology
via the growth factor, matter density and matter power spectrum.
In principle this dependence could be used to constrain all cosmological
parameters
if the intrinsic alignments alone could be observed.
However, we do not believe sufficiently strongly in the linear alignment
model to use it in this way.
We therefore calculate the intrinsic alignment power spectra
$P_{\tilde\gamma^I}(k;\chi)$ and
$P_{\delta,\tilde\gamma^I}$
once, for our fiducial cosmological model, only.
When we make Fisher steps in cosmological parameters we continue to
use these intrinsic alignment power spectra for the fiducial model.
This makes it impossible to constrain parameters such as $\sigma_8$,
and $n_s$ through their effect on the linear alignment model.
This seems sensible to us, and in practice has little effect on our results.
The intrinsic alignment angular power spectra then depend on cosmology
only through the weightings in Eq.s~\ref{eq:powerspec_GI} and~\ref{eq:powerspec_II}.

First we consider the effect on the FoM of unknown intrinsic alignment terms
in the case of perfect photometric redshifts ($\delta_z=0$).
If no II or GI terms are included ${\rm FoM}=38$ for our fiducial
survey.
The upper left panel of Fig.~\ref{fig_fom_nkbin} shows that when the II term is included
we can actually get an increase in FoM if sufficiently little
freedom is given in the intrinsic alignment model!
This is because the II angular power spectra (Eq.~\ref{eq:powerspec_II})
do depend on cosmology
even though we have fixed $P_{\tilde\gamma^I}(k;\chi)$ to be independent
of cosmology.
Because we have here assumed perfect photometric redshifts then the
autocorrelation power spectra measure very well the II angular power spectra
separately from the GG angular power spectra.
Features in the II angular power spectra subtend different angular
scales on the sky and thus constrain cosmology through the standard ruler
effect.
The size of this effect will depend on the fiducial model used for example,
if it is a power law with wavenumber then it would not help to constrain cosmology.
This constraint weakens as more free parameters are used in the II
power spectra. Adding more bins in wavenumber ($x$-axis of panel)
and redshift (upper to lower lines) both degrade the constraints by
a similar but small amount.

The lower left panel of Fig.~\ref{fig_fom_nkbin} shows that in practice
even if $P_{\tilde\gamma^I}(k;\chi)$ has few free parameters it will
not help to constrain cosmology better than if intrinsic alignments
were negligible.
This is because with realistic photometric redshifts it is not possible
to adequately disentangle the GG and II contributions to the shear
power spectra. However the amount of degradation is still relatively
small and flat with increasing number of bins in wavenumber.

The upper and lower middle panels of
Fig.~\ref{fig_fom_nkbin}
show the result of marginalising over an unknown GI power spectrum
in the absence of any II contribution.
The degradation is more severe than for the case with II alone.
We attribute this to the greater similarity between GG and GI
angular power spectra (e.g. Fig~\ref{fig_clbin_fid}).
Therefore it is more difficult to disentangle the two contributions.
Again, the degradation is most sensitive to the number of redshift
bins $m$.
For the most flexible model we consider the degradation is about
a factor of two independent of the photometric redshift error.

Including both II and GI terms at the same time degrades the dark
energy constraints yet further.
With zero photometric redshift errors
the GG+II result is essentially the same as the GG
alone result.
We note that also the result for GG+GI+II is
almost
the same as that for GG+GI.
Thus the II parameters are fitted adequately, presumably using the
information from the autocorrelation power spectra.
However when more realistic photometric redshifts are considered
we see that the II term does cause a degradation.

It is not physical to allow complete freedom in the intrinsic
alignment power spectra. There are various physical processes
that cause the power spectra to vary with scale and redshift.
For example, the processes that tidally align galaxies with
neigboring mass clumps~\citep{hiratas04} may evolve differently to those
which align galaxy light and mass, and thus cause GI correlations~\citep{bridlea07}.
It is very unlikely that there are so many physical processes
that the power spectra could, for example, oscillate in the
redshift direction 5 times between redshift 3 and 0.
Similarly it is hard to think of physical mechanisms to cause
ten changes in power law slope in the k direction.
Our most flexible model has 52 free parameters for the GI
underlying power spectrum and the same again for the II power spectrum.
Considering this large number of free parameters it is perhaps
surprising that the cosmological parameter constraints are not
weakened still further.
We take as our default model for the remainder of the paper
$n=5$, $m=1$, which means we have freedom in 5 bins in the
wavenumber direction as illustrated in Fig.~\ref{fig_clbin_wiggles}
and
the same
power law in the redshift direction for each wavenumber bin.
We also show some results from our maximally free model with
$n=10$ and $m=5$.

We repeat all the calculations using the linear matter power
spectrum in the linear alignment model (Eq.~\ref{eq:emode}
and Eq.~\ref{eq:cross}), shown by the dashed lines in Fig.~\ref{fig_fom_nkbin}.
We see that most of our results are qualitatively similar,
although the FoMs are generally lower.
We attribute this to the difficulty of constraining the parameters
of the intrinsic alignment model from a lower signal (the
linear matter power spectrum is lower than the non-linear matter
power spectrum on small scales).

\begin{figure*}[t]
\center
\epsfig{file=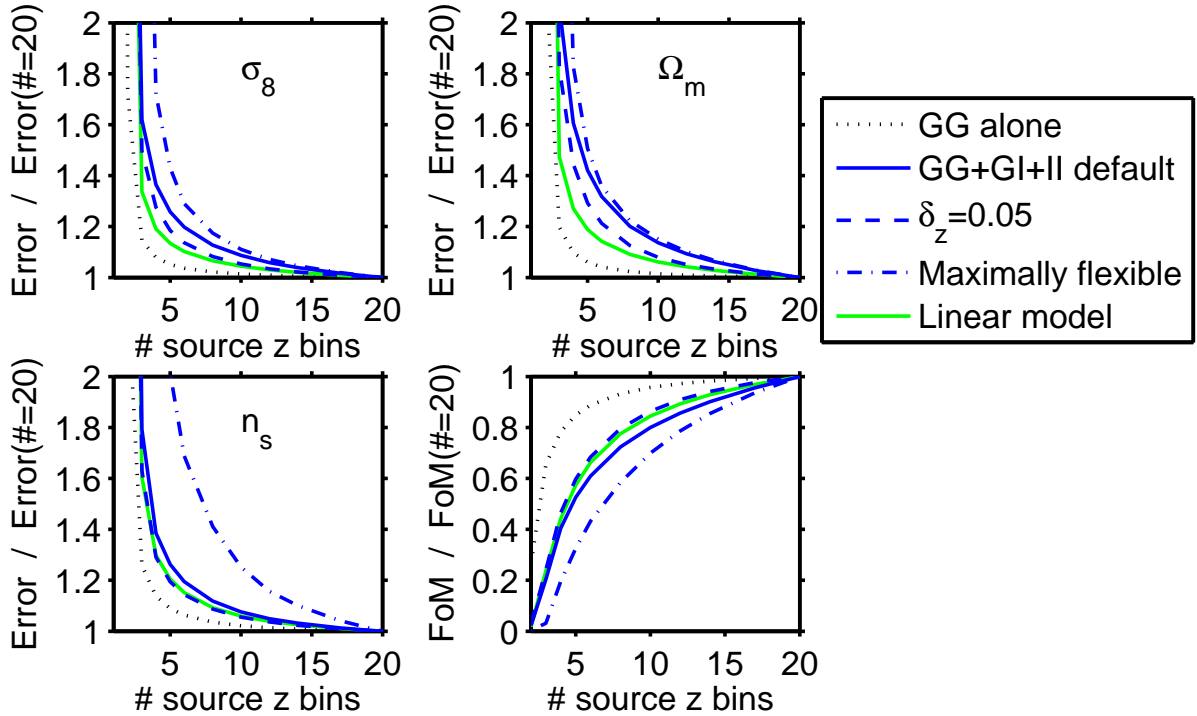,width=16cm}
\caption{
Dependence of cosmological parameter uncertainties on the number
of redshift bins into which the source galaxies are subdivided.
We normalise the performance relative to that for 20 source redshift
bins, to show the fraction of the maximum information that is obtained.
For absolute dark energy Figure of Merit values see the previous figure.
We find that intrinsic alignments require more redshift bins to achieve the
maximum potential.
\emph{Top left:}
Uncertainty divided by minimum uncertainty (at 20 lensed galaxy redshift bins)
for the
fluctuation normalization parameter $\sigma_8$.
\emph{Top right:} Same for the matter density parameter $\Omega_m$.
\emph{Bottom left:} Same for the power spectrum slope parameter $n_s$.
\emph{Bottom right:} Dark energy Figure of Merit, divided
by the maximum Figure of Merit (for 20 lensed galaxy redshift bins).
\emph{Dotted lines:} Results if intrinsic alignments did not exist.
\emph{Solid lines:} Results for our default intrinsic alignment model.
\emph{Dashed lines:} Results using a more realistic photometric
redshift error of $\delta_z=0.05$ (all other lines assume
$\delta_z=0$), default intrinsic alignment flexibility.
\emph{Dot-dashed lines:} A more flexible intrinsic alignment model.
\emph{Light solid lines:} Using the linear theory matter power spectrum
in the linear alignment model instead of the non-linear theory matter
power spectrum, default intrinsic alignment flexibility.
}
\label{fig_all_nzbin}
\end{figure*}

\section{Photometric Redshift Quality}
\label{sec:photoz}

Photometric redshifts exploit spectral features to use
images taken with broad wavelength filters to estimate
redshifts and hence make a three dimensional map of the universe.
We follow~\cite{amarar07} in using two parameters to describe photometric
redshift quality. The first quantifies nearby scatter about the
true spectroscopic redshift $\delta_z = \sigma_z / (1+z)$, where
$\sigma_z$ is the width of the Gaussian contribution centered on the
true redshift.
The second allows a fraction  $f_{\rm cat}$ of catastrophic outliers
to exist at a distance $\pm \Delta_z$ from the central Gaussian distribution.
Throughout we set $\Delta_z=1$ for simplicity.
See the appendix of~\cite{amarar07} and~\cite{abdallaea07} for more details.

\begin{figure*}[t]
\center
\epsfig{file=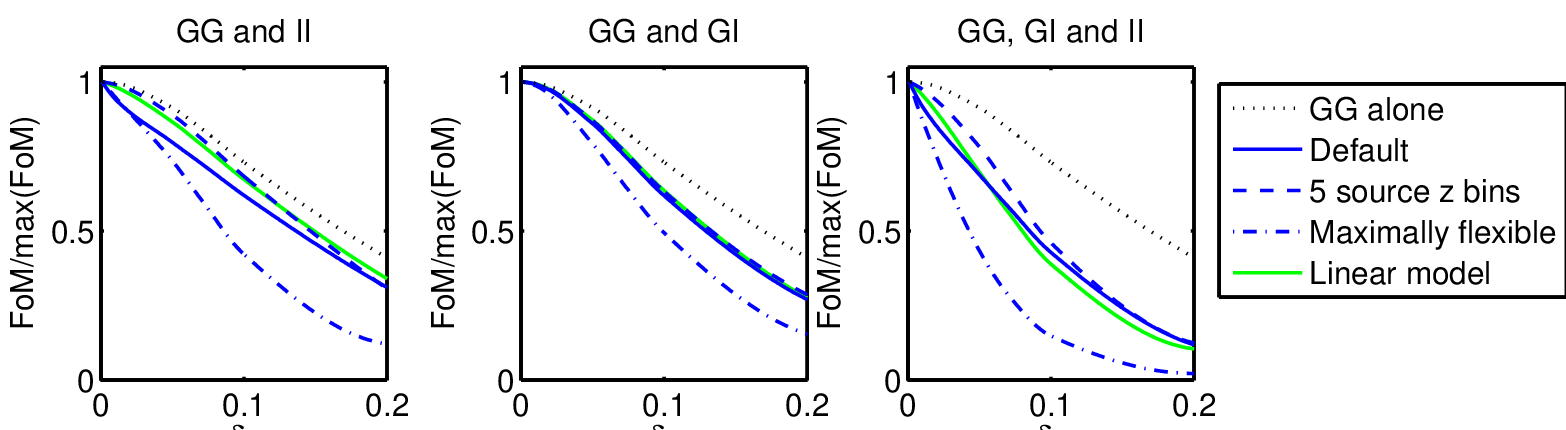,width=18cm}
\caption{
Relative dark energy Figure of Merit as a function of photometric redshift
uncertainty parameter $\delta_z = \sigma_z /(1+z)$.
The FoM degrades more rapidly with decreasing photometric
redshift quality when intrinsic alignments are considered.
\emph{Left panel}: Including just the II term with the usual lensing terms
i.e. assuming that the GI contribution is zero, and known to be so.
\emph{Middle panel}: Including just the GI term (with GG)
i.e. assuming that the II contribution is zero, and known to be so.
\emph{Right panel}: Both II and GI terms included.
\emph{Dotted lines}: Results using just lensing terms alone
(therefore these are the same in every figure),
i.e. assuming that the II and GI terms are zero, and known to be so.
\emph{Solid lines:} Using our default intrinsic alignment model
parametrization.
\emph{Dashed lines:} Using 5 source galaxy photometric redshift bins
(the other lines use 20 bins), default intrinsic alignment model.
\emph{Dot-dashed lines:} A more flexible intrinsic alignment model
($n=10$, $m=5$).
\emph{Light solid lines:} Using the linear theory matter power spectrum
in the linear alignment model instead of the non-linear theory matter
power spectrum, default intrinsic alignment flexibility.
}
\label{fig_fom_deltaz}
\end{figure*}

\subsection{Number of redshift bins}

We first consider the effect of the number of redshift bins used for the
source galaxies on cosmological parameter uncertainties.
For this investigation we assume perfect photometric redshifts by default.
It has been noted by several authors
for the shear signal (GG) alone
(e.g. \cite{hu99,simonks04,mahh06,amarar07,jainct07})
that
although a small amount of tomographic information greatly improves
parameter constraints, continuing to subdivide into further redshift
bins beyond $\sim3$ does not significantly reduce
uncertainties further.
This can be seen by the dotted line in Fig.~\ref{fig_all_nzbin}.
Using about 3 redshift bins brings parameter constraints to within
about 20 per cent of the best obtainable.

However on including intrinsic alignments we expect to need more
information to constrain more parameters.
The amount that the constraints on cosmological parameters are
weakened will depend on the number of free parameters in our
intrinsic alignment modeling.
We start by considering 5 bins in wavenumber
plus an overall
amplitude and redshift power law (7 parameters for each of II and GI).
This corresponds to the circles marked in Fig.~\ref{fig_fom_nkbin}.

The solid line in Fig.~\ref{fig_all_nzbin} shows that
to reduce uncertainties to 20 per cent of the minimum we now need
about 6, 7 or 5 redshift bins for $\sigma_8$, $\Omega_m$ and
$n_s$ respectively. We need 10 redshift bins to bring the
FoM to 80 per cent of its maximum value.
The maximum number of redshift bins we have used is 20, but
if we were to
use more then the numbers of redshift bins required would be larger.
Thus about twice as many (or more) redshift bins are required
to extract the best possible results from the data for the
parameters $\sigma_8$, $\Omega_m$ and $n_s$ twice as many bins
are required, whereas three times as many are needed for the FoM.

For the above we have assumed perfect photometric redshifts.
We now consider a photometric redshift error of $\delta_z=0.05$.
Therefore at the median survey redshift the photometric redshift
uncertainty is $\delta_z (1+z) \sim 0.1$
so there is potential for effectively about
20 redshift bins between $0<z<2$.
The non-zero photometric redshift error
has negligible effect on the lensing alone (GG) curve
and so we do not plot it.
This is in keeping with the fact that a small number of redshift bins
are sufficient to extract most of the information.
For the cases in which we have marginalized over intrinsic
alignment parameters then the non-zero photometric redshift error simply
causes the curves to converge to the
best possible FoM at a smaller number of redshift bins.
This is because the best possible FoM is limited by the photometric
redshift quality, and in effect sets a maximum number of redshift
bins, thus increasing the best obtainable parameter uncertainty.
As an example we show the result for our default intrinsic alignment model
by the dashed lines in Fig.~\ref{fig_all_nzbin}.

We return to perfect photometric redshifts to check the effect
of our choice of parametrization of the intrinsic alignments.
We considered reducing the number of wavenumber bins from 5 to 2
(while keeping just one redshift bin). This made negligible difference
compared to our default parametrization (solid line) so we do not show it in the Figure.
We next switch to our maximally flexible intrinsic alignment model ($n=10$, $m=5$).
The dot-dashed lines in Fig.~\ref{fig_all_nzbin} show how much
the required number of redshift bins is increased.
We now need 8, 9, 10, 12 bins to reach 20 per cent of the best
uncertainty for $\sigma_8$, $\Omega_m$, $n_s$ and the FoM respectively,
a factor of 3 or 4 more than that required by lensing alone.

We repeated all the above results using a different fiducial model
($\sigma_8=0.9$, $\Omega_m=0.3$) and found that these relative
results changed very little, even though the parameter constraints
are tighter for this model with a larger signal.
We see that the results are similar if using the linear matter power
spectrum in the linear alignment model (light solid line in Fig.~\ref{fig_all_nzbin}).
The photometric redshift requirements are less stringent but this is partly
because the constraints are already more degraded (see Fig.~\ref{fig_fom_nkbin}).

\subsection{Photometric redshift error}

Although it is qualitatively helpful to think about the number of
photometric redshift bins, in practice we will have a limited photometric
redshift accuracy and will use as many redshift bins as
computationally possible
(or a fully three dimensional analysis~\cite{heavenskt06}).
We therefore repackage our results in terms
of the photometric redshift accuracy parameter $\delta_z$.
We also investigate the relative importance of the inclusion
of the II and GI terms, but focus in on the FoM for dark energy.

The photometric redshift uncertainties expected from future
surveys depend critically on the number of observing filters.
Using real data with approximately four optical observing bands
a typical scatter $\sigma_z$ is about $0.05$ to $0.1$
\citep{csabaiea03,collisterl04,padmanabhanea05,ilbertea06,abdallaea07}.

Considering the lensing terms alone, we find again the usual
result that the constraints are not very sensitive to
photometric redshift quality (dotted lines in
Fig.~\ref{fig_fom_deltaz}).

Ignoring intrinsic-shear correlations for now, we see in
the left hand panel of Fig.~\ref{fig_fom_deltaz} that
the II term always prefers better photometric redshifts,
when the number of redshift bins is large
i.e. the solid line does not flatten at low $\delta_z$.
This makes sense because the correlation length for
intrinsic alignments
($\sim 10 $ Mpc)
is small compared to the photometric
redshift errors.
For instance, at $z=1$, a comoving distance of 10\,Mpc corresponds to
a $\delta_z=0.004$, and a $\delta_z=0.05$ is 125 Mpc.

Some works propose to remove the II contamination by
removing physically close pairs
\citep{kings02,heymansh03,takadaw04}, instead of
marginalising over a model.
It is clear that for this technique the quality of the
photometric redshifts is paramount if a large number of pairs
are not to be rejected.

To take only a 20 per cent reduction in the FoM (relative
to the best possible) the lensing alone allow
$\delta_z\sim0.08$, whereas to remove the II term of intrinsic
alignments requires twice the precision ($\delta_z\sim0.04$).
The left hand panel of Fig.~\ref{fig_fom_deltaz} shows that
this result is relatively independent of the model used
for intrinsic alignments (solid and dot-dashed lines
are similar at low $\delta_z$).
A very flexible model does very badly for poor photometric
redshift information.
As expected, the number of redshift bins is most important
at high photometric redshift accuracy (low $\delta_z$).

Considering just the GI term (and ignoring the II), the middle
panel of Fig.~\ref{fig_fom_deltaz} shows roughly similar results
to the left hand panel (II but no GI).
The curves are flatter at high photometric redshift accuracy
(smaller $\delta_z$)
which can be explained by the inclusion of the broad lensing
kernel in the GI redshift dependence.
The number of tomographic
redshift bins is not crucial for removing the GI term from
the GG (the solid, 20 bins, and dashed, 5 bins, lines are
very similar).
However the photometric redshift quality is still important.
A twenty per cent reduction in relative FoM requires approximately
30 per cent smaller redshift uncertainties than if no intrinsic
alignments are considered.

Including both the GI and II terms gives a bigger hit on
parameter accuracy than only one term alone.
This is not surprising since we have significantly increased
the number of free parameters, without adding any more data.
The results look qualitatively like those for GG+II alone,
but place even more stringent constraints on photometric
redshift quality.
To obtain 80 per cent of the best possible FoM, the
photometric redshift errors now need to be a factor of 3
to 4 better than if considering lensing alone, depending
on the number of parameters considered in the intrinsic
alignment analysis.

Slicing in the opposite direction, if the photometric
redshift quality is about $\delta_z\sim0.05$, the
lensing constraints are degraded by only 10 per cent,
relative to that for perfect photometric redshifts.
Whereas additionally considering II, GI or GI+II terms
degrades the FoM by 25, 20 and $\sim40$ per cent, relative
to that with perfect photometric redshifts.

For all the combinations, we find the results are not
very sensitive to the number of wavenumber bins used in the II and GI parameterization (n=5 and
n=20 bins give essentially the same curves, whether m=1 or
m=5 redshift bins for the II and GI parametrization are
used).
The light solid line in Fig.~\ref{fig_fom_deltaz} shows the
result using the linear matter power spectrum in the linear alignment
model. Again we see that the requirements are similar but slightly
less stringent.
We repeated the default calculations using an outlier fraction
$f_{\rm cat}=0.1$ but find the results to be negligibly different,
i.e. all the FoMs are degraded by the same factor on changing
$\delta_z$.

We repeated the calculations shown in Fig.~\ref{fig_fom_deltaz} but varying
$f_{\rm cat}$ instead of $\delta_z$ on the x-axis of the figure.
However we found that there was
little difference between the different lines.
All dropped to 80 per cent of the maximum FoM at $f_{\rm cat}\sim0.1$.
When intrinsic alignments (GG+GI+II) were included $f_{\rm cat}$
needed to be 10 per cent smaller than when intrinsic alignments
were assumed to be negligible (GG alone).
When fewer redshift bins were used the requirements on $f_{\rm cat}$
were weaker (5 redshift bins instead of 20 allowed the required
$f_{\rm cat}$ to increase by 30 per cent).
This shows that it is useful to carry
out an analysis with more than 5 redshift bins to make maximum
use of the information available.

\section{Effect of Priors}
\label{sec:priors}

\begin{figure}[t]
\center
\epsfig{file=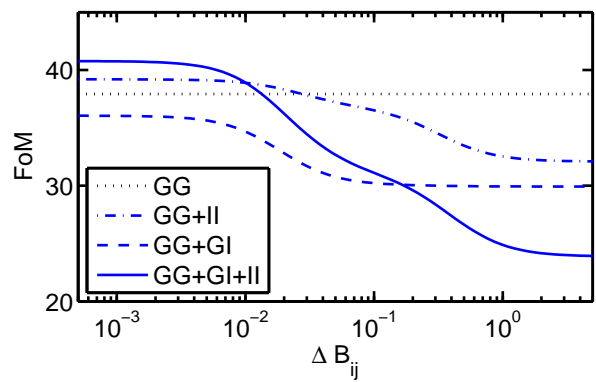,width=8cm}
\caption{
The effect on the dark energy figure of merit as priors on the
intrinsic alignment bin parameter $B_{ij}$ are decreased.
\emph{Dotted line:} Results if intrinsic alignments did not exist.
\emph{Dot-dashed line:} Results if GI did not exist.
\emph{Dashed line:} Results if II did not exist.
\emph{Solid line:} Results using our fiducial intrinsic alignment model
including all contributions.
}
\label{fig_fom_priorB}
\end{figure}

So far we have placed very wide priors on the parameters of the intrinsic
alignment model, and assumed that the photometric redshift distributions
are perfectly known. We now consider narrowing the priors on intrinsic
alignment parameters and allowing uncertainty in the lensed galaxy redshift distributions.
Throughout the below we place the lensed galaxies in 20 photometric redshift
bins, assume default photometric redshift parameters $\delta_z=0.05$,
$f_{\rm cat}=0$ and default intrinsic alignment model with 5 bins in
wavenumber, and
the same
power law in redshift for each wavenumber
($n=5$, $m=1$).

\subsection{Priors on Intrinsic Alignment Model}

We now investigate the impact of
tighter constraints on the intrinsic alignment modeling, that could
come from observations like those of ~\cite{mandelbaumhisb06,hirataea07}
or advances in theoretical predictions~\citep[e.g.][]{heymanswhvv06,vandenboschea02}.
We have previously been using a very wide Gaussian prior on each
bin coefficient $B_{ij}$ of width $\Delta B_{ij}=10$
for numerical stability.
Fig.~\ref{fig_fom_priorB} shows how the FoM increases
as this value is decreased.
We can see that the wide prior we used for the earlier sections
was sufficiently wide that the self-calibration regime was reached.
We also see that if the priors on the IA model are sufficiently
tight then the constraints on cosmology
(GG+GI+II) are tighter than if intrinsic alignments were negligible (GG).

As in Fig~\ref{fig_fom_nkbin}, GG+GI gives a lower FoM than GG+II.
In addition the transition to the self-calibration regime occurs at smaller
$\Delta B_{ij}$ for GG+GI than for GG+II. This means that more precise
information is required to alleviate the cosmological parameter
degradation, compared to GG+II.
The exact transition to self-calibration will in general depend
on the number of parameters being fitted, however for our default
model with 7 parameters for each of II and GI, priors on the
perturbing parameters $B_{ij}$ need to be less than 10 per cent
to help improve cosmological parameters constraints.
We repeated the calculation keeping the overall amplitude and power law
parameters fixed but the qualitative picture remains the same.

\subsection{Priors on redshift distributions}

In practice the redshift distributions of the lensed galaxies cannot
be known perfectly because we only have imperfect photometric
redshifts.
So far we have dealt with the accuracy properties of each single
photometric redshift estimate, in terms of a scatter parameter
and a catastrophic outlier parameter.
For cosmic shear it is generally much more important to have accurate knowledge
of these parameters than for these parameters to be small.
It is difficult to know the values of these parameters accurately
since they depend on uncertain modeling
of galaxy formation~\citep{abdallaea07} and/or large test samples
for which spectroscopic redshifts are known.

Here, we discuss our knowledge of these parameters by considering our
uncertainty on them, and place Gaussian priors on their values.
This follows previous work~\citep{huterertbj06,mahh06,jainct07,amarar07},
which has also discussed the numbers of spectroscopic redshifts needed to obtain
these priors.
We revert back to wide priors on the intrinsic alignment parameters
$\Delta B_{ij}=10$.

We divide the redshift range $0<z<3$ up into 30 photometric redshift prior bins
such that an equal number of galaxies fall into each bin.
We apply a shift of $z_{{\rm bias} (i)}$ to all the spectroscopic
redshifts in bin number $i$.
We assign the scatter for bin $i$ to be $\delta_{z (i)}(1+z)$.
We then allow these 30 parameters to vary in the Fisher matrix and
apply priors of width $\Delta \delta_{z}$ and $\Delta z_{\rm bias}$
which are the same for every photometric redshift prior bin.
We then create the joint probability distribution $P(z_s,z_p)$
and make cuts in photometric redshift space, as before.

This is the same as in~\cite{mahh06} except we vary $\delta_z$
instead of $\sigma_z=\delta_z(1+z)$ and using binning with equal
numbers of galaxies in each bin
instead of even spacing in redshift.
The broad results are similar
but more accentuated when we switch to the~\cite{mahh06} method.

\begin{figure}[t]
\center
\epsfig{file=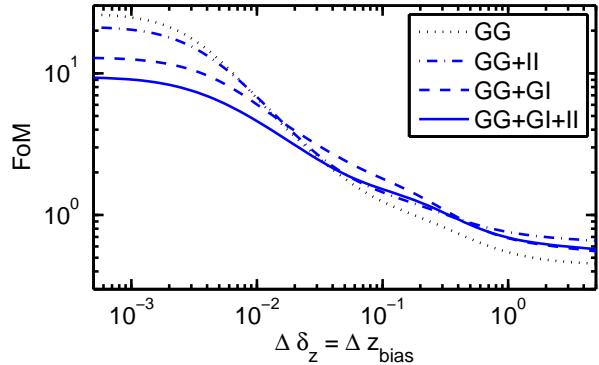,width=8cm}
\caption{The effect of loosening our priors on photometric redshift
parameters. Previously we effectively assumed zero for the
uncertainty on photometric the redshift scatter $\Delta \delta_z (1+z)$
and on the uncertainty in photometric redshift bias $\Delta z_{\rm bias}$.
Lines as in previous figure.}
\label{fig_zpriors}
\end{figure}

The results in Fig.~\ref{fig_zpriors} show that in terms
of absolute FoM, tighter priors are required when intrinsic alignments
are considered, if we are to obtain the same FoM as in their absence.
For example, to obtain an FoM of 10 extremely tight priors are
required if considering GG+GI+II, whereas a wider prior could be used
if intrinsic alignments did not exist (GG).
However, if we are considering the \emph{relative} figure of merit then
the results are reversed.
To obtain 80 per cent of the best possible FoM, the presence of
intrinsic alignments allows us to loosen our requirements on the
priors, and thus allow smaller spectroscopic redshift training sets.
Note that this is because the best possible FoM is lower when we
take into account intrinsic alignments.

If we had particularly poor prior knowledge of
photometric redshift properties, then the existence of intrinsic
alignments actually helps to calibrate the uncertain parameters.
This is particularly true for the intrinsic-intrinsic term.
Note that this all assumes our default photometric redshift fiducial
model in which $\delta_z=0.05$ and $f_{\rm cat}=0$.
Therefore if photometric redshifts were good, but we did not know it,
then intrinsic alignments would help.

For our example with 30 photometric redshift parameter bins
the FoM with intrinsic alignments included
(GG+GI+II) is better than that if intrinsic alignments did not
exist for
$\Delta z_{\rm bias} = \Delta \delta_z > 0.03$.
However, a considerable fraction of the FoM has been lost by this
stage so we hope not to be in this regime.
Following~\cite{mahh06} we would already have a prior tighter than this
if we had a number
$N^{\mu}_{\rm spec} = (\sigma_z /\Delta z_{\rm bias})^2  = ( (\delta_z (1+z)) /\Delta z_{\rm bias} )^2 \sim 10$
spectroscopic redshifts in each
photometric redshift parameter bin.
It is likely we will already have more spectroscopic redshifts than this, but %S
perhaps this feature might become useful for very deep surveys,
for which spectroscopy will be harder to obtain, or as a cross-check.

\section{Conclusions}

We have compared our fiducial model to current observations and an
alternative model and found it to be in good agreement.
We use the linear alignment model as our base, and investigate
inserting the non-linear matter power spectrum, as a way of
taking into account small scale correlations
(as suggested in \cite{hirataea07}).
We find that this matches better to the HRH* model for
intrinsic-intrinsic (II) alignments, and matches better to the
GI detections of~\cite{mandelbaumhisb06,hirataea07}.

We allow this model freedom as a function of scale and redshift
since we will not know the intrinsic alignment contributions
perfectly in time for future dark energy surveys, due to the
dependence on baryonic physics.
We choose as our baseline model one which has five free parameters
as a function of scale, and
the same
power law function of
redshift for each scale.
We allow the option of adding more redshift variation, and more or
less parameters as a function of wavenumber.
We assume the II and GI perturbations about the linear alignment
model are independent of each other.

We find that if little freedom is given to the intrinsic
alignment modeling then the constraints on dark energy
parameters can actually be
very slightly
improved, relative to those
obtained if intrinsic alignments do not exist.
Essentially the intrinsic alignments are providing another
cosmological probe.

The shear-intrinsic correlations are more subtle since
they contain the lensing efficiency function.
It seems plausible that cross correlations between
a given redshift bin and the various other redshift bins
will allow simultaneous constraints on the GI term
and cosmology.
This is not so dissimilar to the lensing geometric test
\citep{jaint03,huj04,bernsteinj04}, in which the mass
of halos is jointly constrained with cosmology due to
the dependence of the shear signal with redshift.
However, on allowing realistic freedom in the intrinsic alignment
modeling and using more realistic photometric redshift
uncertainties the potential is lost, and dark energy
constraints are degraded by about 50 per cent.

We show how the number of source galaxy redshift bins
affects cosmological parameter constraints with and
without intrinsic alignments.
We find that on considering intrinsic alignments the
number of redshift bins must be doubled or tripled to
obtain the full information available.

We investigate in more detail the effect of photometric
redshift quality on constraints. We consider how constraints
are degraded with respect to the case in which perfect
photometric redshifts are available.
We confirm that intrinsic alignments put tighter constraints
on photometric redshift quality than lensing alone.
We find roughly equal contributions from each of the II and
GI terms.
The photometric redshift error required is approximately
three times smaller that using lensing alone, to extract the same
fraction of the available information.
It would therefore be most unwise to plan future
gravitational lensing experiments to capitalize on the result
that lensing does not require precise redshift information.
The color tomography method of~\cite{jainct07} could likely
still be applied if enough observing bands were used.

A significant catastrophic outlier fraction
would be bad news for
methods for removing the intrinsic-intrinsic correlations
by excluding pairs of galaxies from the analysis.
Statistically it may be known that there is a ten per cent
chance the true redshift is far from that measured.
Deciding whether or not to remove the pair is then more unclear,
and contamination will leak in.
Modeling the intrinsic-intrinsic correlations means that
statistical information on the redshift distribution is sufficient
i.e. we can use the number as a function of redshift for each bin rather than
a precise redshift of each galaxy.
However we do need to use an intrinsic alignment
model that can encapsulate the true intrinsic alignment contribution.

\cite{mandelbaumhisb06} suggested excluding
brightest cluster galaxies
(BCGs; which often pass the cuts used to
select
luminous red galaxies,
LRGs) from samples used to measure
cosmic shear, in order to minimize the intrinsic terms. However, these galaxies do have smaller
photometric redshift errors and hence would be useful to include in 3-D lensing studies,
as well as in the constraint of cosmological parameters that we have considered here.
Perhaps simultaneously fitting for the intrinsic terms as well as for the cosmic shear signal as a function of galaxy type (plus relative redshift and scale), rather than excluding BCGs would be possible.
In this paper we have ignored the fact that the intrinsic alignment
parameters depend on galaxy type.

Obtaining additional information on the intrinsic alignments
from other sources would be extremely valuable, since
the degradation of cosmological parameter constraints is large
on using a very flexible model for intrinsic alignments.
This could come from better numerical simulations
or from observations of the galaxy-shear and galaxy-galaxy
cross correlation functions such as those by
\cite{mandelbaumhisb06,hirataea07} or as suggested by
\cite{huj04}.
We find that priors of around 10 per cent would be helpful.

Cosmic shear produces only E-mode (curl-free) distortions, except on
sub-arcminute scales where source clustering induces a B-mode signal
\citep{schneidervmm02}. The presence of B-mode distortions would indicate
that residual systematics are present, either from incomplete correction of
galaxy ellipticities for distortions (associated with the atmosphere or
telescope) or from intrinsic galaxy alignments. The magnitudes of the B-mode
power spectra of II and GI are highly model dependent - thus, if one can be
sure that the data have been accurately corrected for other systematics,
this signal will place stringent constaints on which physical models of
intrinsic alignment are supported by the data.

To summarise our findings:
\begin{itemize}
\item{
When constraining the dark energy equation of state using a cosmic shear survey, it is {\em essential} to account for intrinsic-intrinsic galaxy alignments (II correlations) and the cross-term between intrinsic ellipticity and cosmic shear (GI correlations). Neglecting these correlations biases the dark energy equation of state by $\sim$50\%.}
\item{
We considered a range of flexibility in the intrinsic alignment
modelling and found a wide variation in the impact on the degradation in
the dark energy figure of merit. It will be necessary for future
observational and theoretical studies to inform this choice when making
future dark energy constraints.}
\item{In general when II and GI correlations are accounted for in cosmological parameter estimation, the number of tomographic redshift bins must be a factor of at least two more than if only lensing correlations   are present, to obtain 80\% of the available information.}
\item{Concerning the dark energy figure of merit, the presence of II and GI correlations places a more stringent requirement on the accuracy of photometric redshifts - a factor of $\sim 3$ -  compared with the case where only cosmic shear is present.}
\end{itemize}

\section*{Acknowledgements}

We thank
Gary Bernstein,
Tom Kitching,
Masahiro Takada,
Filipe Abdalla,
Chris Blake,
Patrick Simon,
Adam Amara,
Alexandre Refregier,
Antony Lewis,
Bhuvnesh Jain
and
Ofer Lahav
for helpful discussions.
SLB and LJK acknowledge support from Royal Society University Research Fellowships.

%\bibliography{../../slb}

\end{document}